\documentclass[aps,preprint]{revtex4}
\usepackage{graphicx}

\begin{document}

\title{Ising Model on Edge-Dual of Random Networks}

\author{A. Ramezanpour}
\email{ramzanpour@mehr.sharif.edu}
 \affiliation{Department of Physics, Sharif University of
Technology, P.O.Box 11365-9161, Tehran, Iran}

\date{\today}

\begin{abstract}
We consider Ising model on edge-dual of uncorrelated random
networks with arbitrary degree distribution. These networks have a
finite clustering in the thermodynamic limit. High and low
temperature expansions of Ising model on the edge-dual of random
networks are derived. A detailed comparison of the critical
behavior of Ising model on scale free random networks and their
edge-dual is presented.

\end{abstract}

\maketitle

\section{Introduction}\label{1}
It is evident that a variety of natural and artificial systems can
be described in terms of complex networks, in which the nodes
represent typical units and the edges represent interactions
between pairs of units \cite{ab,d,n1}. Clearly, identifying
structural and universal features of these networks is the first
step in understanding the behavior of these systems
\cite{ws,ba,asbs,ajb,n2,fdj}. Intensive research in recent years
has revealed peculiar properties of complex networks which were
unexpected in the conventional graph theory \cite{b}. Among these
one can refer to the scale free behavior of degree distribution
\cite{ba}, $P(k)$, where degree denotes the number of nearest
neighbors of a node. From another point of view one is interested
in the effect of structural properties of complex network on the
collective behavior of systems living on these networks
\cite{bw,g,sv,ak,ja,mb,khk,h}. Percolation and Ising model (or in
general Potts model) are typical examples of statistical mechanics
which have intensively been studied on uncorrelated random
networks with given degree distributions
\cite{dgm,nsw,cnsw,gdm,lvvz,cah,cabh,bp,vm,vw,dgm1}. By
uncorrelated random network we mean those in which the degree of
two neighbors are independent random variables . These
 networks are identified only by a degree distribution, $P(k)$, and
have the maximum possible entropy. The locally tree-like nature of
these networks provides a good condition to apply the recurrence
relations to study the collective behavior of interesting
models\cite{bax,dgm,dgm1}. It is seen that, depending on the level
at which the higher moments of $P(k)$ become infinite, one
encounters different critical behaviors that could be derived from
a landau-Ginzburg theory\cite{gdm}. This in turn reflects the mean
field nature of these behaviors.\\

\begin{figure}
  \centering
\includegraphics[width=12cm]{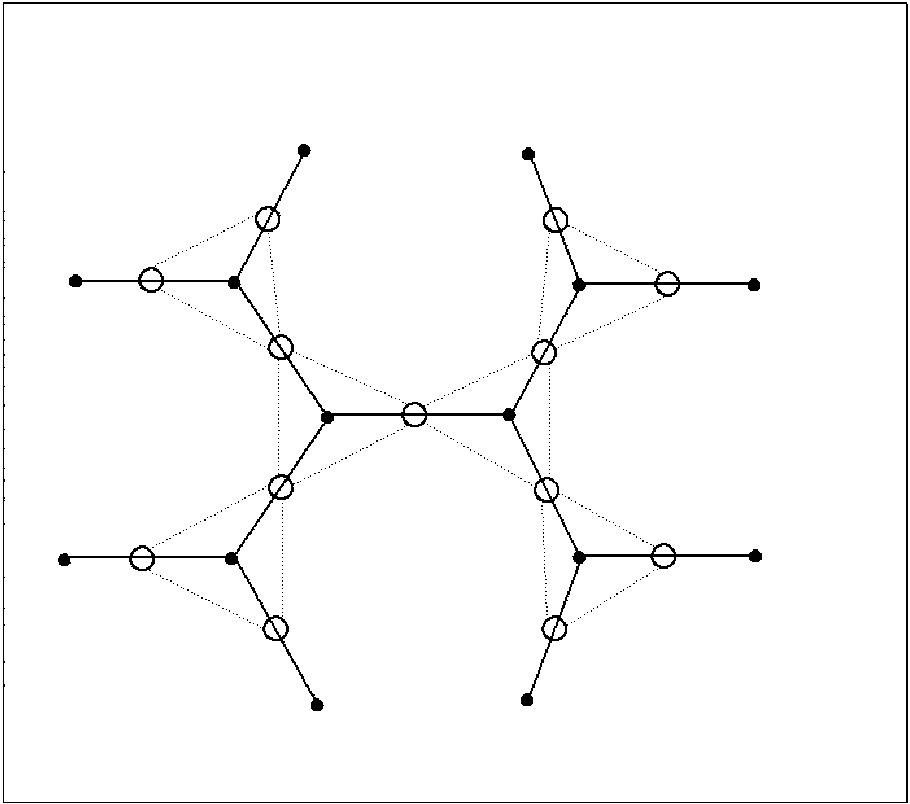}
    \caption{Some part of the edge-dual of a Bethe lattice with $z=3$. Empty circles and
    dotted lines denote the nodes and the edges of the edge-dual network respectively.}\label{f1}
    \end{figure}

In this paper we are going to study the Ising model on the
edge-dual of uncorrelated random networks with a given degree
distribution. These kinds of networks have already been introduced
in the context of graph theory\cite{b}. Given a network $G$, its
edge-dual $\tilde{G}$ can be constructed as follows, see also
figure(\ref{f1}): one puts a node in place of each edge of $G$ and
connects each pair of these new nodes if they are emanating from
the same node of $G$. \\ Such networks have been useful among
other things for the study of maximum matching problem \cite{zy}
as well as topological phase transitions in random
networks\cite{pdfv}.\\
The interesting point about these networks is that they have
generally a large degree of clustering even when the underlying
network is tree-like. Due to this high clustering direct study of
structural properties of such networks or of physical models
defined on them is usually very difficult. However one can use
this duality to adopt the techniques used in the context of random
networks (e.g. the generating function formalism \cite{nsw,cnsw})
and obtain interesting results for such networks \cite{rkm}. For
example it was shown in \cite{rkm} that the edge-dual of a random
scale free network with $P(k)\propto k^{-\gamma}$ will be a
network whose degree distribution behaves like
$\tilde{P}(k)\propto k^{-\tilde{\gamma}}$ for large degrees where
$\tilde{\gamma}=\gamma-1$.\\

Our basic result is depicted in table \ref{tab}, where we have
compared the critical behavior of Ising model on scale free
networks and their edge-dual networks. On the way to this basic
result we have also studied as a preliminary step the Ising model
on the edge-dual of Bethe lattices. We have also developed a
systematic high and low temperature expansion for the Ising model
on edge-dual networks for arbitrary degree distributions.\\
Moreover as a byproduct we have also shown that there is a simple
relation between the partition function of an Ising model on a
tree in which each spin interacts with its nearest and next
nearest neighbors and the partition function of an Ising model on
its edge-dual with only nearest neighbor interactions but in the
presence of magnetic field. \\

The paper is organized as follows. In section \ref{2} a
relationship between the Ising model on a tree and on its
edge-dual is derived. High and low temperature expansions of the
partition function of the Ising model on the edge-dual of a random
network are given in section \ref{3}. In section \ref{4} we use
recurrence method for the study of Ising model on the edge-dual of
Bethe lattices. The same method is applied in section \ref{5} to
study the critical behavior of Ising model on the edge-dual of
scale free networks. The conclusions are presented in section
\ref{6}.

\section{A Relation Between Ising Model on a Tree And Its Edge-Dual Network}\label{2}
\begin{figure}
  \centering
\includegraphics[width=12cm]{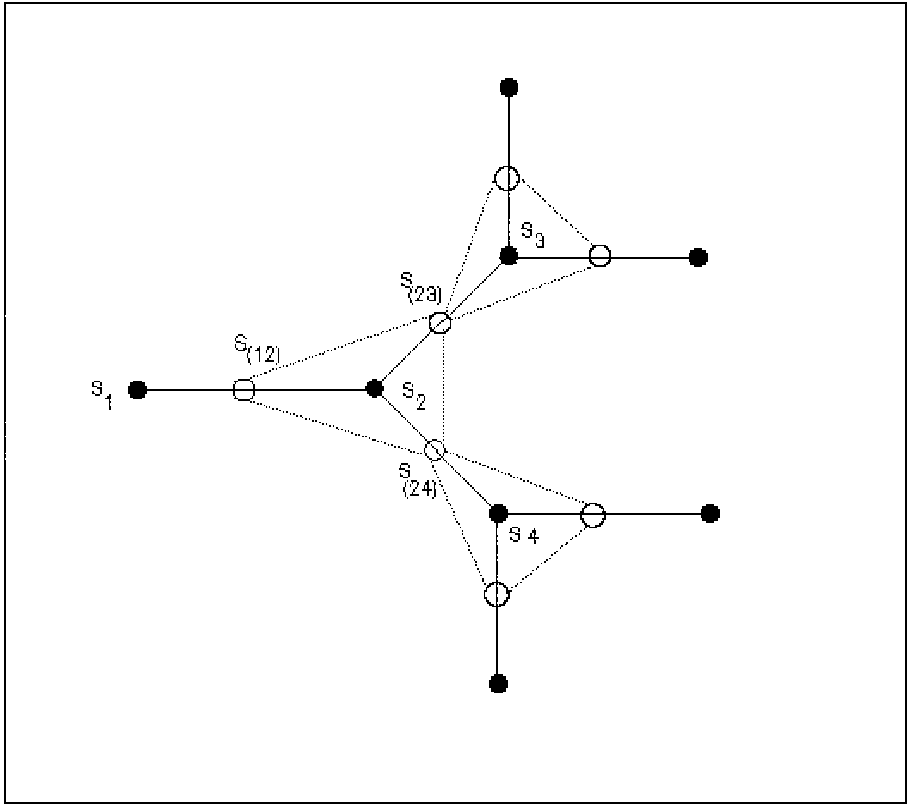}
    \caption{To each edge $(ij)$ we can assign a spin
    which is defined as the product of spins at the end point nodes of that edge i.e. $S_{(ij)}:=S_iS_j$.
    }\label{f2}
    \end{figure}

Let us consider an Ising model (with values of spin taking only
$\pm 1$) on a tree graph $G$ with nearest and next nearest
neighbor interactions of strength $J_1$ and $J_2$ in the absence
of magnetic field. The hamiltonian is

\begin{equation}\label{Htree}
E=-J_1\sum_{<ij>_1}S_iS_j - J_2\sum_{<ij>_2}S_iS_j,
\end{equation}

where $<ij>_1$ and $<ij>_2$ denote the nearest and next nearest
neighbors respectively. For any given configuration of spins we
can assign a unique configuration of spin variables (again taking
values $\pm 1 $) to the edges of the graph: any edge which
connects two nodes having the variables $ S_i $ and $ S_j $ is
assigned a value $S_{(ij)}:=S_iS_j$, see figure (\ref{f2}). On the
other hand for any configuration of spins on the edges,
${S_{(ij)}}$, there are two possible configuration of spins on the
nodes, which are obtained by flipping all the spins $ S_i $ on the
graph. Therefore there is a two to one correspondence between the
spin configurations on the nodes of the graph and the edges of the
graph.

Now if we write the above hamiltonian in terms of spins of edges
we get

\begin{equation}\label{Htreed}
E=-J_1\sum_{<ij>_1}S_{(ij)} - J_2\sum_{<ij>_2}S_{(ik)}S_{(kj)},
\end{equation}

where $k$ is the common nearest neighbor of nodes $i$ and $j$. But
this is the hamiltonian of an Ising model on $\tilde{G}$, the
edge-dual of the tree, with nearest neighbor interactions of
strength $\tilde{J_1}=J_2$ in presence of a magnetic field of
magnitude $\tilde{h}=J_1$.

\begin{equation}\label{Htd}
\tilde{E}=-\tilde{h}\sum_{\tilde{i}}S_{\tilde{i}} -
\tilde{J_1}\sum_{<\tilde{i}\tilde{j}>_1}S_{\tilde{i}}S_{\tilde{j}},
\end{equation}

where $ \tilde{i} $ and $\tilde{j} $ now denote the nodes of the
edge-dual graph. Taking into account the relation between
configurations mentioned above we obtain

\begin{equation}\label{ZZd}
Z(J_1,J_2,h=0,T)=2\tilde{Z}(\tilde{J_1}=J_2,\tilde{J_2}=0,\tilde{h}=J_1,T),
\end{equation}

in which $Z$ and $\tilde{Z}$ are respectively the partition
functions of the Ising model on $G$ and $\tilde{G}$ and $T$
denotes the temperature. We should stress that the above relation
is true only for tree graphs, since the presence of loops in $G$
puts constraints on the values of spins which are assigned to the
nodes of $\tilde{G}$. That is for any loop in $G$, the product of
spins on its edges (equivalently the nodes of $\tilde{G}$) should
be $ +1$. Taking into account all these constraints makes the
calculation of the partition function very difficult in the
general case.\\

In the thermodynamic limit the two models have the same free
energy density, that is $f=\tilde{f}$, where $f:=\frac{-T\ln
Z}{N}$ and $\tilde{f}:=\frac{-T\ln \tilde{Z}}{\tilde{N}}$ are
respectively the free energy of the Ising model on tree and its
edge-dual. Here we have set the Boltsmann constant equal to one.
Note that for a tree graph, $\tilde{N}=N-1$. Moreover, it is easy
to see that the average magnetization of a spin in $\tilde{G}$

\begin{equation}\label{m}
\tilde{m}=-\frac{\partial \tilde{f}}{\partial \tilde{h}}
\end{equation}

is equal to the average correlation of two neighboring spins in
$G$

\begin{equation}\label{ss}
<S_iS_j>=\frac{-\partial f}{\partial J_1}.
\end{equation}

Knowing the average magnetization of spins in $\tilde{G}$ as a
function of magnetic field, we can write the free energy density
from equation (\ref{m}) as follows \cite{bax}

\begin{equation}\label{fd}
\tilde{f}(\tilde{J_1},\tilde{h},T)=\int_{\tilde{h}}^{\infty}(\tilde{m}(h')-1)dh'
-\tilde{J_1} - \tilde{h}.
\end{equation}

Differentiation of the right hand side with respect to $\tilde{h}$
correctly gives the magnetization $ \tilde{m}$ and the integration
constant $-\tilde{J_1}-\tilde{h}$ can be understood from the fact
that for very large magnetic fields when the first integral
vanishes, the partition function is dominated by the
configurations where all the spins are up, and hence the free
energy per site is given by $-\tilde{J_1}-\tilde{h}$.\\
Obviously if we replace $\tilde{J_1}$ with $J_2$ and $\tilde{h}$
with $J_1$ we obtain $f$, the free energy per site of Ising model
on $G$ with nearest and next nearest neighbor interactions. In
this way any nonanalytic behavior of $\tilde{f}$ will appear in
$f$ too.

\section{High And Low Temperature Expansions of Ising Model on Edge-Dual Networks}\label{3}
Consider a random network which we denote by $ G$. We assume that
the probability of each node having degree $k$ is equal to $
P(k)$. Furthermore we assume that there is no correlations between
the degrees of adjacent neighbors. The average number of nearest
neighbors will be denoted by $n_1:=<k>=\sum_k kP(k)$. Degree
distribution of nearest neighbors\cite{cnsw} is given by
$\Pi(k)=\frac{kP(k)}{<k>}$. Thus the average number of next
nearest neighbors will be
$n_2:=<k>\sum_k (k-1)\Pi(k)=<k^2>-<k>$.\\
The edge-dual of such a network is denoted by $ \tilde{G}$.
Correspondingly every quantity pertaining to the dual network
$\tilde{G}$ will be designated by a tilde sign.\\
In the following we will study the Ising model with only nearest
neighbor interactions on $\tilde{G}$. The partition function is

\begin{equation}\label{Z0}
\tilde{Z}=\sum_{\{S_{\tilde{i}}\}}e^{\tilde{K}\sum_{<\tilde{i}\tilde{j}>}S_{\tilde{i}}S_{\tilde{j}}},
\end{equation}

where the sum in the exponential is over the nearest neighbors on
the edge-dual of $G$. Here we have used the notation
$\tilde{K}:=\frac{\tilde{J}}{T}$.

\subsection{High Temperature Expansion}\label{3.1}
The above partition function can be written in a form appropriate
for a high temperature expansion\cite{bax}. To this end we write
the exponential in the form

\begin{equation}\label{exp}
e^{\tilde{K}\sum_{<\tilde{i}\tilde{j}>}S_{\tilde{i}}S_{\tilde{j}}}=
\cosh^{\tilde{L}}(\tilde{K})\prod_{<\tilde{i}\tilde{j}>}(1+S_{\tilde{i}}S_{\tilde{j}}\tanh(\tilde{K})),
\end{equation}

where $\tilde{L}:=\frac{Nn_2}{2}$ is the number of edges in the
$\tilde{G}$. Inserting this in equation (\ref{Z0}) and expanding
the product we get a series of terms each corresponding to a
subgraph of $\tilde{G}$. Summing over spin configuration only
terms which represent closed loops will survive \cite{bax} and we
arrive at the following expression for the partition function

\begin{equation}\label{Z1}
\tilde{Z}=\cosh^{\tilde{L}}(\tilde{K})2^{\tilde{N}}\sum_{{\rm{closed\
loops\ c}}} \tilde{\eta}^{\tilde{L}(c)},
\end{equation}

where $\tilde{\eta} = \tanh (\tilde{K}) $,
$\tilde{N}=\frac{Nn_1}{2}$ is the number of nodes of $\tilde{G}$
and $\tilde{L}(c)$ is the perimeter (the number of edges) of the
closed loop $c$.

Clearly at high temperatures the first and the second terms
corresponding to triangles and squares need be kept in the
expansion.

\begin{equation}\label{w2}
\tilde{Z}=\cosh^{\tilde{L}}(\tilde{K})2^{\tilde{N}}(
\tilde{N}_{\triangle}\tilde{\eta}^3+\tilde{N}_{\diamondsuit}\tilde{\eta}^4+
\ldots.)
\end{equation}

The number of triangles in $ \tilde{G}$ has two parts: first, each
triangle of $G$ appears as a triangle in $\tilde{G}$ too.
Secondly, by definition of the edge-dual network, every triple of
edges emanating from the same node in $G$ make a triangle in
$\tilde{G}$. The number of the latter types of triangles is given
by the number of distinct choices of three edges of a node, summed
over the nodes of $G$. Thus

\begin{equation}\label{triangl}
\tilde{N}_{\triangle}=N_{\triangle}+\sum_i (\begin{array}{c}
  k_i \\
  3 \\
\end{array}),
\end{equation}

in which $k_i$ is the degree of node $i$ in $G$. Since in an
uncorrelated random network the number of triangles is a finite
quantity in the thermodynamic limit \cite{cpv}, we can neglect the
first term compared with the second one which has an infinite
contribution in this limit. The same argument is applicable to the
case of squares so we can approximate these numbers by

\begin{eqnarray}\label{trisqu}
\tilde{N}_{\triangle} \approx \sum_i (\begin{array}{c}
  k_i \\
  3 \\
\end{array})=\frac{N}{3!}\sum_k k(k-1)(k-2)P(k),\\\nonumber
\tilde{N}_{\diamondsuit} \approx \sum_i (\begin{array}{c}
  k_i \\
  4 \\
\end{array})=\frac{N}{4!}\sum_k k(k-1)(k-2)(k-3)P(k).
\end{eqnarray}

Note that these relations become exact in the case of tree
structures even for finite $N$.

\subsection{Low Temperature Expansion}\label{3.2}
We now return to equation (\ref{Z0}), the original relation for
the partition function. Note that we can rewrite it as

\begin{equation}\label{Z0'}
\tilde{Z}=\sum_{\{S_{(ij)}\}}e^{\frac{\tilde{K}}{2}\sum_i((\sum_{j|i}S_{(ij)})^2-k_i)},
\end{equation}

where the first sum in the exponential is over nodes of $G$ and
the second is over nearest neighbors of that node. Let us write
$\tilde{Z}$ in a simpler way

\begin{equation}\label{Z1'}
\tilde{Z}=e^{-\tilde{K}L}\sum_{\{S_{(ij)}\}}e^{\frac{\tilde{K}}{2}\sum_i
m_i^2},
\end{equation}

where $L$ is the number of edges in $G$ and
$m_i:=\sum_{j|i}S_{(ij)}$. Using the following identity

\begin{equation}\label{I}
e^{\frac{\tilde{K}}{2}m_i^2}=\frac{1}{\sqrt{\frac{2\pi}{\tilde{K}}}}\int
dx_ie^{-\frac{\tilde{K}}{2}({x_i}^2 +2 m_ix_i)},
\end{equation}

we find

\begin{equation}\label{Z2'}
\tilde{Z}=e^{-\tilde{K}L}(\frac{\tilde{K}}{2\pi})^{N/2}\sum_{\{S_{(ij)}\}}\int
Dx e^{-\frac{\tilde{K}}{2}\sum_i( x_i^2 + 2x_i m_i)},
\end{equation}

where $Dx = \prod_i dx_i $ and $i$ runs over all the nodes of $ G
$. We can now perform the sum over the spin configurations in the
integrand. To this end we note in view of the definition of $ m_i
$

\begin{equation}\label{sum1}
 \sum_{\{S_{(ij)}\}} e^{-\tilde{K}\sum_i x_i m_i } = \sum_{\{S_{(ij)}\}}
 e^{-\tilde{K}\sum_{<ij>}(x_i+x_j)S_{(ij)}}
 \end{equation}

where $\sum_{<ij>}$ sums over all the links of $G$. The sum can be
transformed to

\begin{eqnarray}\label{sum2}
 & &\prod_{<ij>}(e^{-\tilde{K}(x_i + x_j)} + e^{\tilde{K}(x_i +
 x_j)})\cr &=& e^{\tilde{K}\sum_{<ij>}(x_i+x_j)}\prod_{<ij>}(1+e^{-2\tilde{K}(x_i +
 x_j)})\cr &=& e^{\tilde{K}\sum_i k_i x_i}\prod_{<ij>}(1+e^{-2\tilde{K}(x_i +
 x_j)})
 \end{eqnarray}

 Putting all these together we find

\begin{equation}\label{Z3'}
\tilde{Z}=e^{-\tilde{K}L}(\frac{\tilde{K}}{2\pi})^{N/2}\int Dx
e^{-\frac{\tilde{K}}{2} \sum_i (x_i^2-2k_i x_i)}
\prod_{<ij>}(1+e^{-2\tilde{K}(x_i+x_j)}).
\end{equation}

The product $\prod_{<ij>}(1+e^{-2\tilde{K}(x_i+x_j)})$ can now be
expanded as a series of terms each corresponding to a subgraph $g$
of $G$.

For any node $i$ of a  the graph $G$, a factor $ e^{-2Kz_ix_i}$
should be taken into account in which $ z_i $ is the degree of
that node in the subgraph. If a node $i$  dose not belong to the
subgraph, $z_i = 0 $. Any subgraph determines uniquely the
sequence of integers $ \{z_i; i = 1, \ldots N\} $. Note that $ z_i
\leq k_i \ \ \  \forall i $. For each such sequence the integral
can be easily calculated yielding

\begin{equation}\label{Z4'}
\tilde{Z}=e^{-\tilde{K}L+\frac{\tilde{K}}{2}\sum_i
k_i^2}\sum_{{\rm g}} e^{-2\tilde{K}\sum_i z_i(k_i-z_i)}.
\end{equation}

It is the central result of this subsection which can be used for
a low temperature expansion of Ising model on $\tilde{G}$. This
formula incidentally shows that each subgraph $g$ and its
complement (the graph obtained when one removes all the links of $
g $ from $G$) give the same contribution to the partition
function. \\

At very low temperatures, $\tilde{K}\longrightarrow \infty$, only
the empty graph for which all $ z_i = 0 $ and its complement for
which all $ z_i = k_i $ contribute yielding

\begin{equation}\label{Z5'} \tilde{Z_0}=2e^{-\tilde{K}L+
\frac{\tilde{K}}{2}\sum_i {k_i}^2}=2e^{-\tilde{K}L+
\frac{\tilde{K}}{2}N\sum_k
k^2P(k)}=2e^{-\tilde{K}\frac{N}{2}\langle k\rangle +
\frac{\tilde{K}}{2}N\langle k^2\rangle}
\end{equation}

resulting in a free energy per site equal to

\begin{equation}\label{f0}
    \tilde{f_0} =
-\frac{\tilde{J}}{2}(\langle k^2\rangle - \langle k \rangle)
\end{equation}

where we have used the relation $
\tilde{K}=\frac{\tilde{J}}{T}$.\\
The next to leading order term comes from subgraphs which have
only one link, (we multiply their contribution by 2 to account for
their complements). This will give

\begin{equation}\label{Z6'}
    \tilde{Z} = \tilde{Z}_0 (1+ \sum_{<ij>} e^{-2\tilde{K}(k_i + k_j-2)}+\cdots )
\end{equation}

where the sum is over all the links of $G$. This can be written as
follows

\begin{equation}\label{Z7'}
    \tilde{Z} = \tilde{Z}_0 ( 1 + N\frac{\langle
    k\rangle}{2}e^{4\tilde{K}}\ll
    e^{-2\tilde{K}(k+k')}\gg+\cdots)
\end{equation}

where $\ll\gg$ denotes the average with respect to the two point
function $ P(k,k')$, the probability of two nodes of degrees $ k $
and $ k'$ to be neighbors. For uncorrelated networks one has $
P(k,k') = (2-\delta_{k,k'})\Pi(k)\Pi(k')$. This procedure can be
followed
for higher order contributions.\\

\section{Ising Model on Edge-Dual of Bethe Lattices}\label{4}

The Bethe Lattice \cite{bax} is defined as a regular network where
all nodes have the same degree $z$. Let us consider Ising spins on
the edges of this network and let them interact with an external
magnetic field $ \tilde{h}$ and with each other if their
corresponding edges are incident on the same node of the Bethe
lattice. We can write the average magnetization of a spin lying on
an arbitrary edge using the following recurrence
relation\cite{bax}

\begin{equation}\label{mb}
\tilde{m}=\frac{e^{\tilde{H}}
g_+^2(0)-e^{-\tilde{H}}g_-^2(0)}{e^{\tilde{H}}g_+^2(0)+e^{-\tilde{H}}g_-^2(0)},
\end{equation}

where $\tilde{H}:=\frac{\tilde{h}}{T}$ and $g_+(0)$ and $g_-(0)$
are respectively the partition functions for the system of spins
on one side of the central spin when it is up or down. That is

\begin{equation}\label{g}
g_S(0)=\sum_{\{S_R\}}e^{\tilde{K}\sum_{\tilde{i}|S}SS_{\tilde{i}}+
\tilde{K}\sum_{<\tilde{i}\tilde{j}>}S_{\tilde{i}}S_{\tilde{j}}+\tilde{H}\sum_{\tilde{i}}S_{\tilde{i}}},
\end{equation}

\begin{figure}
  \centering
\includegraphics[width=12cm]{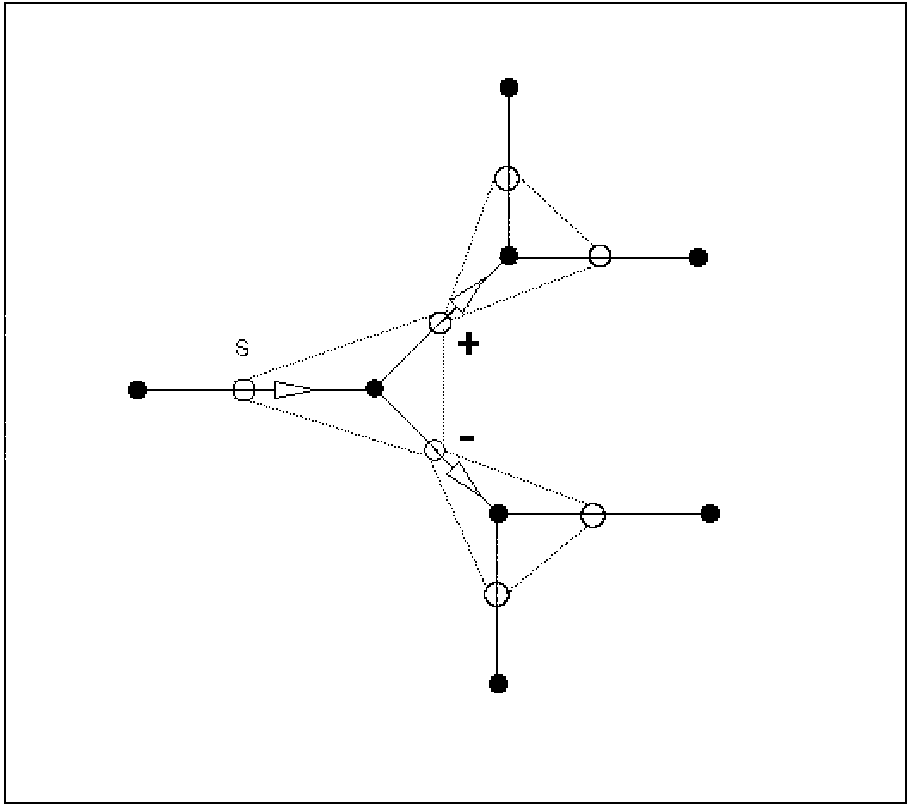}
    \caption{The partition function of spins on the branch stems from $S$
    can recursively be written in terms of the partition functions of branches now
    stem from nearest neighbors of $S$.}\label{f3}
    \end{figure}

where as before $\tilde{K}:=\frac{\tilde{J}}{T}$. Note that in
this partition function only spins on one side of $S$ (for
instance the right hand side spins, denoted by $\{S_R\}$) appear.
Similar to $ g_S(0) $ one can define a partition function $ g_S(l)
$ which gives the partition function of the branch of the lattice
which stems from a node at layer $l$ where the value of its spin
has been fixed to $S$. These partition functions can be related to
each other recursively as follows, see also figure (\ref{f3}): in
the right hand of $S$ there are $z-1$ spins which interact with
$S$ and with each other. We can write $g_S(l)$ as a sum over
different configurations of these spins. For each configuration we
will have a term proportional to $g_+^r(l+1)g_-^{z-1-r}(l+1)$
where $r$ will be the number of up spins in such a configuration.
Moreover we have to consider another factor which takes into
account the Boltzmann factor associated to this configuration of
spins. Energy of a configuration in which $r$ of these spins are
up is the sum of three parts, a part given by their interaction
with the external magnetic field equal to $-\tilde{h}(2r-z+1)$, a
part from their interactions with spin $S$ equal to
$-\tilde{J}(Sr-S(z-1-r))$ and finally a part given by interactions
between themselves equal to
$-\tilde{J}(\frac{r(r-1)}{2}+\frac{(z-1-r)(z-1-r-1)}{2}-r(z-1-r))$.
Summing up the above arguments we arrive at

\begin{equation}\label{gl}
g_S(l)=\sum_{r=0}^{z-1}(\begin{array}{c}
  z-1 \\
  r \\
\end{array})
e^{\tilde{H}(2r-z+1)+\tilde{K}(S(2r-z+1)-2r(z-1-r)+\frac{(z-1)(z-2)}{2})}g_+^r(l+1)
g_-^{z-1-r}(l+1).
\end{equation}

Returning to equation (\ref{mb}), magnetization of the central
spin can be rewritten in a simpler form

\begin{equation}\label{mb1}
\tilde{m}=\frac{e^{2\tilde{H}} -x_0^2}{e^{2\tilde{H}}+x_0^2}=
\frac{e^{2\tilde{H}} -e^{-2y_0}}{e^{2\tilde{H}}+e^{-2y_0}},
\end{equation}

where we have defined $x_l:=\frac{g_-(l)}{g_+(l)}=:e^{-y_l}$. When
the magnetic field is positive we have $g_-(l)<g_+(l)$ thus $y_l$
is a positive quantity which plays a role similar to the magnetic
field and thus can be interpreted as the local field experienced
by a spin at distance $l$ from the central spin $S$. Now using
equation (\ref{gl}), the recurrence relation for $y_l$ reads

\begin{equation}\label{yl}
y_l=-ln\left(\frac{\sum_r(\begin{array}{c}
  z-1 \\
  r \\
\end{array})
e^{(\tilde{2H}+y_{l+1})r+\tilde{K}(z-1-2r-2r(z-1-r))}}{\sum_r(\begin{array}{c}
  z-1 \\
  r \\
\end{array})
e^{(\tilde{2H}+y_{l+1})r+\tilde{K}(2r-z+1-2r(z-1-r))}}\right).
\end{equation}

Setting $\tilde{H}=0$ and starting from distant ($l\gg 1$) spins
with $y \ll 1$ one could  obtain the values of $y$ for deeper
spins in a step by step manner using the above relation until one
arrives at $y_0$. Equation (\ref{mb1}) tells us that we will have
magnetization in this case only if $y_0$ is different from zero.
It is evident that a stable nonzero solution for $y_0$ is possible
only when the right hand side of the recurrence relation for
$y_l$( when plotted versus $y_{l+1}$) has a slope greater than or
equal to $1$. The equality will provide the critical temperature
of the system which turns out to be given by

\begin{equation}\label{tc}
\frac{\sum_r r(\begin{array}{c}
  z-1 \\
  r \\
\end{array})
e^{-\tilde{K}_c2r(z-1-r)}2\sinh(\tilde{K}_c(2r-z+1))}{\sum_r(\begin{array}{c}
  z-1 \\
  r \\
\end{array})
e^{\tilde{K}_c(z-1-2r(z-r))}}=1.
\end{equation}

\begin{figure}
  \centering
\includegraphics[width=10cm,angle=-90]{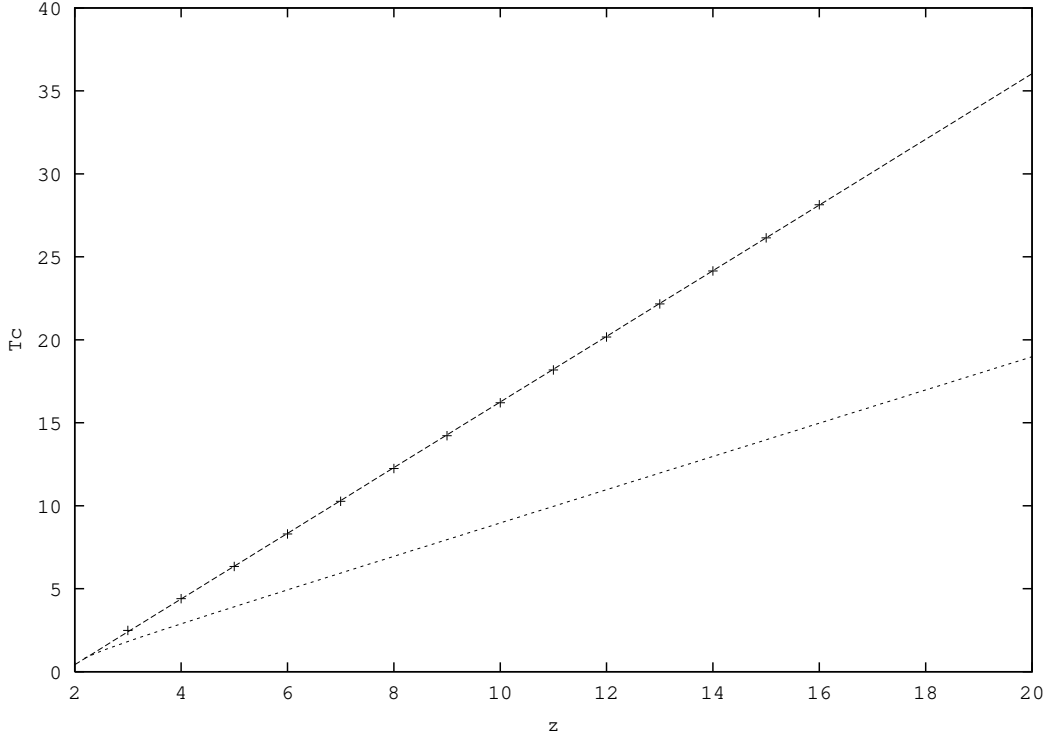}
    \caption{$T_c$( in units of $J$) for the Ising model
    on Bethe lattices(lower curve) and their edge-dual(upper curve).}\label{f4}
    \end{figure}

Unfortunately it is not possible to derive a closed relation for
$\tilde{K}_c$. In figure (\ref{f4}) we have computed this quantity
numerically and compared it with the corresponding quantity in the
Bethe lattice itself. In the latter case the critical temperature
reads \cite{bax}

\begin{equation}\label{tcb}
\tanh{K_c}=\frac{1}{z-1}.
\end{equation}

As figure (\ref{f4}) shows $\tilde{T}_c$ is much grater than $T_c$
which is as expected due to the larger number of interactions in
the edge-dual network. In the figure we also show a linear fit,
$\tilde{T}_c=a_0+a_1z$, to the numerical data for $\tilde{T}_c$
with $a_0=-3.53 \pm 0.02$ and $a_1=1.97 \pm 0.002$. As long as the
critical behavior of the system is concerned we expect to see a
standard mean field behavior as in the usual Ising model in
spatial dimensions greater than $d_c=4$. We will further discuss
these issues in the next section.

\section{Ising Model on Edge-Dual of Random Networks}\label{5}
In this section we generalize the results of the previous section
to the case of Ising model on the edge-dual of random networks
with a given degree distribution $P(k)$. In this case a spin on
the edge of such a random network will encounter $k_1-1$ and
$k_2-1$ nearest neighbors at its right and left hand sides
respectively. These numbers are random variables given by the
degree distribution of nearest neighbors in the random network
$\Pi(k)$.

\subsection{General Arguments}\label{5.1}
Along the lines of section \ref{4} we can write the magnetization
of a spin on an edge of random network with end point nodes of
degrees $k_1$ and $k_2$ as

\begin{equation}\label{mkk'}
\tilde{m}_{k_1k_2}=\frac{e^{2\tilde{H}}-e^{-y_0(k_1)-y_0(k_2)}}{e^{2\tilde{H}}+e^{-y_0(k_1)-y_0(k_2)}},
\end{equation}

where now $y$'s are random variables depending on distance and
degree of end point nodes. Obviously magnetization of an arbitrary
spin is given by

\begin{equation}\label{mr}
\tilde{m}=\sum_{k_1,k_2}\Pi(k_1)\Pi(k_2)\frac{e^{2\tilde{H}}-e^{-y_0(k_1)-y_0(k_2)}}
{e^{2\tilde{H}}+e^{-y_0(k_1)-y_0(k_2)}}.
\end{equation}

As before here we used the notation $x_l(k):=e^{-y_l(k)}$ and
$x_l(k):=\frac{g_-(l;k)}{g_+(l;k)}$ where $g_S(l;k)$ is the
partition function for the cluster beyond the spin $S$ at distance
$l$ from the central spin. As in the case of Bethe lattices
(figure (\ref{f3})) these quantities can  be related to
$g_{S'}(l+1;k')$'s by the following relations

\begin{equation}\label{gS}
g_S(l;k)=\sum_{r=0}^{k-1}e^{\tilde{H}(2r-k+1)+\tilde{K}(S(2r-k+1)-2r(k-1-r)+\frac{(k-1)(k-2)}{2})}
\sum_{r/k-1}\prod_{a=1}^r
g_+(l+1;k_a)\prod_{b=r+1}^{k-1}g_-(l+1;k_b),
\end{equation}

where the second sum is over different selections of $r$ distinct
spins from the set of $k-1$ neighboring spins after assigning
indices $1$ to $r$ to them. Thus the relation for $x_l(k)$ gets
the form

\begin{equation}\label{xlk}
x_l(k)=\frac{\sum_{r=0}^{k-1}e^{\tilde{H}2r+\tilde{K}(k-1-2r-2r(k-1-r))}
\sum_{k-1-r/k-1}\prod_{a=1}^{k-1-r} x_{l+1}(k_a)}
{\sum_{r=0}^{k-1}e^{\tilde{H}2r+\tilde{K}(2r-k+1-2r(k-1-r))}
\sum_{k-1-r/k-1}\prod_{a=1}^{k-1-r} x_{l+1}(k_a)},
\end{equation}

or in terms of $y$'s

\begin{equation}\label{ylk}
y_l(k)=-ln(\frac{\sum_{r=0}^{k-1}e^{\tilde{H}2r+\tilde{K}(k-1-2r-2r(k-1-r))}
\sum_{k-1-r/k-1}e^{-\sum_{a=1}^{k-1-r} y_{l+1}(k_a)}}
{\sum_{r=0}^{k-1}e^{\tilde{H}2r+\tilde{K}(2r-k+1-2r(k-1-r))}
\sum_{k-1-r/k-1}e^{-\sum_{a=1}^{k-1-r} y_{l+1}(k_a)}}).
\end{equation}

Let us also derive a relation for the average energy of Ising
model on the edge-dual of random networks in the absence of
magnetic field. First note that we can write this quantity as a
sum over the interaction energy associated to the spins on the
edges emanating from the same node of $G$, that is

\begin{equation}\label{Ee}
\tilde{E}=\sum_ie_i=\sum_i(\sum_{<(ij)(ik)>}-\tilde{J}S_{(ij)}S_{(ik)}).
\end{equation}

Thus the thermodynamic average of the above quantity reads

\begin{equation}\label{aEe}
<\tilde{E}>=N\sum_k P(k)<e_k>,
\end{equation}

where $<e_k>$ is the average energy associated to a node of degree
$k$. We are able to write this quantity in terms of $g$'s by
summing over different configurations of spins on the edges of
such a node. As before we sum over configurations in which $r$
spins out of these $k$ spins are up. For each such configuration
we include an appropriate Boltzmann weight as before.\\
 Denoting
by
$e_k:=-\tilde{J}(\frac{r(r-1)}{2}+\frac{(k-r)(k-r-1)}{2}-r(k-r))$
the interaction energy of these spins we obtain

\begin{equation}\label{ek}
<e_k>=\frac{\sum_{r=0}^k e_k
e^{\tilde{K}(\frac{r(r-1)}{2}+\frac{(k-r)(k-r-1)}{2}-r(k-r))}
\sum_{r/k}\prod_{a=1}^r g_+(0;k_a)\prod_{b=r+1}^{k}g_-(0;k_b)}
{\sum_{r=0}^k
e^{\tilde{K}(\frac{r(r-1)}{2}+\frac{(k-r)(k-r-1)}{2}-r(k-r))}
\sum_{r/k}\prod_{a=1}^r g_+(0;k_a)\prod_{b=r+1}^k g_-(0;k_b)}.
\end{equation}

After some algebra this relation takes the following simpler form
in terms of $y$'s

\begin{equation}\label{eky}
<e_k>=-\tilde{J}\left(\frac{k(k-1)}{2}-\frac{\sum_{r=0}^k 2r(k-r)
e^{-\tilde{K}2r(k-r)} \sum_{k-r/k}e^{-\sum_{a=1}^{k-r}y_0(k_a)}}
{\sum_{r=0}^k e^{-\tilde{K}2r(k-r)}
\sum_{k-r/k}e^{-\sum_{a=1}^{k-r} y_0(k_a)}}\right).
\end{equation}

Equations (\ref{mr}), (\ref{ylk}) and (\ref{eky}) are the exact
relations for  magnetization, effective fields $y$ and energy of
the system.

\subsection{The Effective Medium Approximation}\label{5.2}
In this section we simplify the relations obtained in the previous
subsection using the effective medium approximation
\cite{dgm,dgm1} applied satisfactorily to the study of Ising model
on uncorrelated random networks. It is believed that this
approximation takes in a good way into account the effects of high
degree nodes which play an essential role in determining the
critical behavior of the system specifically in inhomogeneous
network having scale free
degree distribution.\\
To this end we rewrite the relations derived above as if $y$'s are
independent of $k$, the degree of the end pint nodes. This is
achieved if we use the same $g$ for all the spins which are at the
same distance from the central spin. The only explicit dependence
on $k$ enters equation (\ref{ylk}) which must be averaged over
using the degree distribution of nearest neighbors $\Pi(k)$. We
emphasize that this approximation is exact if we expand our
relations for small $y$'s and keeping only the linear term. Note
that we are finally interested in the critical behavior of the
system where $y$'s tend to zero and thus we expect the above
approximation to work well in the critical region. Consequently we
use the following relations to extract the critical behavior of
Ising Model on the edge-dual of an uncorrelated random network;

\begin{equation}\label{mef}
\tilde{m}_=\frac{e^{2\tilde{H}}-e^{-2y_0}}{e^{2\tilde{H}}+e^{-2y_0}},
\end{equation}

for the magnetization,

\begin{equation}\label{ylef}
y_l=-\sum_k \Pi(k)ln\left(\frac{\sum_{r=0}^{k-1}(\begin{array}{c}
  k-1 \\
  r \\
\end{array})
e^{(2\tilde{H}+y_{l+1})r+\tilde{K}(k-1-2r-2r(k-1-r))}}
{\sum_{r=0}^{k-1}(\begin{array}{c}
  k-1 \\
  r \\
\end{array})
e^{(2\tilde{H}+y_{l+1})r+\tilde{K}(2r-k+1-2r(k-1-r))}}\right),
\end{equation}

for the recurrence relations defining $y$'s and

\begin{equation}\label{eef}
<\tilde{E}>=-N\tilde{J}\sum_k
P(k)\left(\frac{k(k-1)}{2}-\frac{\sum_{r=0}^k (\begin{array}{c}
  k \\
  r \\
\end{array}) 2r(k-r)
e^{-\tilde{K}2r(k-r)+y_0r}} {\sum_{r=0}^k
e^{-\tilde{K}2r(k-r)+y_0r}}\right).
\end{equation}

for the average of energy in the absence of magnetic field.\\

At this stage it is instructive  to note that we can obtain the
correlation between the central spin and a spin at distance $l$ by
takin derivative of $\tilde{m}$ with respect to $\tilde{H}_l$, the
magnetic field acting on such a spin. To this end we need also to
label the magnetic fields along with the $y$'s in the above
relations. Indeed in equations (\ref{mef}) and (\ref{ylef}) the
magnetic field has a similar index to that of y. On the other hand
we have

\begin{equation}\label{corr}
\tilde{\chi}_l:=\frac{\partial\tilde{m}}{\partial
\tilde{H}_l}|_{\{\tilde{H}_l=0\}}=\tilde{n}_l \tilde{G}_c(0,l),
\end{equation}

where $\tilde{n}_l=2(\frac{<k^2>-<k>}{<k>})^l$ is the number of
spins at distance $l$ from the central spin in the edge-dual of
random network and $\tilde{G}_c(0,l):=<SS_l>-<S><S_l>$. Note that
susceptibility is given by $\tilde{\chi}=\sum_l \tilde{\chi}_l$
where from (\ref{corr}) and (\ref{mef}) $\tilde{\chi}_l$ reads

\begin{equation}\label{corrc}
\tilde{\chi}_l=\frac{\partial\tilde{m}}{\partial
y_0}\prod_{i=0}^{l-2} (\frac{\partial y_i}{\partial
y_{i+1}})\frac{\partial y_{l-1}}{\partial
\tilde{H}_l}|_{\{\tilde{H}_l=0\}}.
\end{equation}

If spins are deep enough in the network we can take all the $y$'s
equal to each other, so for their derivatives. After making this
approximation we obtain

\begin{equation}\label{corrc1}
\tilde{\chi}_l=\frac{\partial\tilde{m}}{\partial
y_0}(\frac{\partial y_i}{\partial y_{i+1}})^{-1}\frac{\partial
y_{l-1}}{\partial
\tilde{H}_l}e^{-\frac{l}{\tilde{\lambda}}}|_{\{\tilde{H}_l=0\}},
\end{equation}

where

\begin{equation}\label{corrl}
 \tilde{\lambda}:=-\frac{1}{ln(\frac{\partial y_i}{\partial y_{i+1}})}.
\end{equation}

Here the index $i$ is only to distinguish between $y$'s in two
subsequent shells and we will eventually set all the $y$'s equal
to each other. This quantity is determined from the fixed point of
equation (\ref{ylef}). Consequently the length scale
$\tilde{\lambda}$ is determined from equation (\ref{corrl}) and as
expected, it will become infinite in the critical point, that is
when $\frac{\partial y_i}{\partial y_{i+1}}=1$. It is this
critical behavior that gives rise to the critical behavior of
$\tilde{\chi}$. Note however that $\tilde{\lambda}$ is not the
correlation length which is determined from the long distance
behavior of $\tilde{G}_c(0,l)$.

\subsection{The critical behavior}\label{5.3}
Using equation (\ref{ylef}) it is not difficult to see that $y_0$
is nonzero only for temperatures less than $\tilde{T}_c$ which
satisfies

\begin{equation}\label{tcr}
\sum_k \Pi(k)\frac{\sum_{r=0}^{k-1} r(\begin{array}{c}
  k-1 \\
  r \\
\end{array})
e^{-\tilde{K}_c2r(k-1-r)}2\sinh(\tilde{K}_c(2r-k+1))}{\sum_r(\begin{array}{c}
  k-1 \\
  r \\
\end{array})
e^{\tilde{K}_c(k-1-2r(k-r))}}=1.
\end{equation}

It is a simple generalization of equation (\ref{tc}). This
equation tells us that if $\tilde{K}_c \rightarrow 0$, i.e. at
high temperatures, we have $\tilde{K}_c \sim
\frac{<k>}{<k^2>-<k>}$. In other words the critical temperature of
the system becomes infinite only when the second moment of $P(k)$
is infinite. This is the same behavior observed in uncorrelated
random networks \cite{dgm,lvvz}. Thus even the finite value of
clustering of edge-dual networks can not significantly alter the
critical point although $\tilde{T}_c \gg T_c$ as we saw in the
case of Bethe lattices in
section \ref{4}.\\

Now let us limit ourselves to the critical region where $y$,
$\tilde{H}$ and
$\tilde{\tau}:=|\frac{T-\tilde{T}_c}{\tilde{T}_c}|$ are very
small. We want an expansion of $\tilde{m}$, $y$ and $<\tilde{E}>$
in terms of small deviations from the critical point. First note
that if we change the sign of $\tilde{H}$ , then by definition the
sign of $y$ changes too and thus the magnetization is reversed. On
the other hand, energy does not change under this change of sign.
Any expansion of these quantities in terms of $\tilde{H}$ and $y$
must satisfy these symmetries. We summarize these arguments in the
following expansions

\begin{eqnarray}\label{expan}
\tilde{m}\approx \tilde{H}+y \\ \nonumber y\approx
a_1(2\tilde{H}+y)+a_3(2\tilde{H}+y)^3
\\ \nonumber <\tilde{E}>\approx b_0 + b_2 y^2,
\end{eqnarray}

where the coefficients $a_1, a_3, b_0 $  and  $b_2 $, are found to
be

\begin{eqnarray}\label{expanco}
a_1=1+O(<k^4>)\tilde{\tau} \hskip 1cm a_3\sim O(<k^4>)\\ \nonumber
b_0 \sim O(<k^2>) \hskip 1cm b_2 \sim O(<k^4>).
\end{eqnarray}

In this equation the averages are taken with respect to the degree
distribution of random network $P(k)$. To be more specific let us
consider a definite degree distribution, that is the well known
scale free distribution $P(k)\propto k^{-\gamma}$. We consider
several cases depending on the value of $ \gamma $.

\subsubsection{The case $\gamma >5$}
In this case the edge-dual network behaves as a scale free network
\cite{rkm}, with $\tilde{\gamma}=\gamma-1
> 4$. Moreover, all the coefficients appearing in the expansions of
(\ref{expan}) are finite. It is easy to show using equations
(\ref{expan}) and (\ref{expanco}) that $y$ is given by following
relations

\begin{eqnarray}\label{yc1}
y\sim \tilde{\tau}^{\frac{1}{2}} \hskip 1cm \tilde{H}=0\\
\nonumber y \sim \tilde{H}^{\frac{1}{3}} \hskip 1cm \tilde{\tau}=0 \\
\nonumber y\sim \frac{\tilde{H}}{\tilde{\tau}} \hskip 1cm
\tilde{H}\neq 0 , \tilde{\tau} \neq 0.
\end{eqnarray}

The critical behavior of the other quantities can easily be
derived from these relations

\begin{eqnarray}\label{qc1}
\tilde{m}\sim \tilde{\tau}^{\frac{1}{2}},\hskip 0.5cm \delta
\tilde{C}\sim cons.,\hskip 0.5cm
\tilde{\chi} \sim \tilde{\tau}^{-1} \hskip 1cm \tilde{H}=0\\
\nonumber \tilde{m} \sim \tilde{H}^{\frac{1}{3}} \hskip 1cm
\tilde{\tau}=0,
\end{eqnarray}

where $\delta \tilde{C}$ is the change of specific heat through
the critical point. Here we have only shown dependence of
interesting quantities on $\tilde{\tau}$ and $\tilde{H}$. Clearly
these behaviors are those of the standard mean field model seen in
the Ising model in spatial dimensions greater than $d_c=4$. This
behavior is also seen in the case of Ising model on uncorrelated
scale free random networks with $\gamma>5$ \cite{dgm,lvvz}.\\ Note
that due to the finiteness of all the moments of degree
distribution in Bethe lattices, the critical behavior of Ising
model on their edge-dual network also lies in this class.

\subsubsection{The case $\gamma=5$}
Now $\tilde{\gamma}=4$. Some of the coefficients in expansions of
(\ref{expan}) become infinite. To avoid these divergences which
are an artifact of our expansion, we set a cut off for degrees
which is proportional to $\frac{1}{y}$. Indeed in our expansion we
used the fact that $ky\ll 1$ where $k$ is the degree of a nearest
neighbor. Fortunately we are interested in the critical behavior
where $y\rightarrow 0$ and the the above arguments work well in
that region \cite{dgm}.\\
Considering the above arguments, we find

\begin{eqnarray}\label{yc2}
y\sim \tilde{\tau}^{\frac{1}{2}} \hskip 1cm \tilde{H}=0\\
\nonumber y
\sim (\frac{\tilde{H}}{ln(\tilde{H})})^{\frac{1}{3}} \hskip 1cm \tilde{\tau}=0 \\
\nonumber y\sim \frac{\tilde{H}}{\tilde{\tau} ln(\tilde{\tau})}
\hskip 1cm \tilde{H}\neq 0 , \tilde{\tau} \neq 0.
\end{eqnarray}

Thus the interesting quantities behave as

\begin{eqnarray}\label{qc2}
\tilde{m}\sim \tilde{\tau}^{\frac{1}{2}},\hskip 0.5cm \delta
\tilde{C}\sim ln(\tilde{\tau}),\hskip 0.5cm
\tilde{\chi} \sim \frac{1}{\tilde{\tau} ln(\tilde{\tau})} \hskip 1cm \tilde{H}=0\\
\nonumber \tilde{m} \sim
(\frac{\tilde{H}}{ln(\tilde{H})})^{\frac{1}{3}} \hskip 1cm
\tilde{\tau}=0.
\end{eqnarray}

\subsubsection{The case $3<\gamma<5$}
In this case the degree distribution of edge-dual network will
have the exponent $2<\tilde{\gamma}<4$.
 Again we have to take into account the divergences
appearing in the expansion coefficients. From equations
(\ref{expan}) and (\ref{expanco}) we find

\begin{eqnarray}\label{yc3}
y\sim \tilde{\tau}^{\frac{1}{2}}\hskip 1cm \tilde{H}=0\\ \nonumber
y \sim
\tilde{H}^{\frac{8-\gamma}{9}} \hskip 1cm \tilde{\tau}=0 \\
\nonumber y\sim
\frac{\tilde{H}}{\tilde{\tau}^{\frac{\gamma-3}{2}}} \hskip 1cm
\tilde{H}\neq 0 , \tilde{\tau} \neq 0.
\end{eqnarray}

And for the thermodynamic quantities we find

\begin{eqnarray}\label{qc3}
\tilde{m}\sim \tilde{\tau}^{\frac{1}{2}},\hskip 0.5cm \delta
\tilde{C}\sim \tilde{\tau}^{\frac{\gamma-5}{2}},\hskip 0.5cm
\tilde{\chi} \sim \tilde{\tau}^{-\frac{\gamma-3}{2}}  \hskip 1cm \tilde{H}=0\\
\nonumber \tilde{m} \sim \tilde{H}^{\frac{8-\gamma}{9}} \hskip 1cm
\tilde{\tau}=0.
\end{eqnarray}\\

We do not consider the case $\gamma \leq 3$ since in this region
$\tilde{\gamma} \leq 2$ and the average number of neighbors is
infinite in the edge-dual network although it is still finite in
the corresponding random network.\\

The above results show that these critical behaviors are very
different from the ones seen in the uncorrelated scale free
networks \cite{dgm,lvvz}, see table \ref{tab} . For example here
the magnetization always behaves like the standard mean field case
, $\tilde{m} \sim \tilde{\tau}^{\frac{1}{2}} $, but in
uncorrelated scale free networks this behavior is only seen for
$\gamma > 5$ where all quantities are of the standard mean field
type \cite{dgm,lvvz}.

\begin{table}

\begin{center}
\begin{tabular}{|c|c|c|c|}
  \hline

               & Magnetization     & Specific heat       & Susceptibility \\
  \hline
  $\gamma > 5$ & $ \tilde{\tau}^{\frac{1}{2}} \left(\tau^{\frac{1}{2}}\right)$ \vline $\ \ \tilde{H}^{\frac{1}{3}}$
               &  cons.$\left(cons.\right)$ & $ \tilde{\tau}^{-1}\left(\tau^{-1}\right)$ \\

  \hline
  $\gamma = 5$ & $ \tilde{\tau}^{\frac{1}{2}}\left((\frac{\tau}{ln(\tau)})^{\frac{1}{2}}\right)$ \vline
  $\ \
(\frac{\tilde{H}}{ln(\tilde{H})})^{\frac{1}{3}}$ & $
ln(\tilde{\tau})\left(\frac{1}{ln(\tau)}\right)$ & $ \frac{1}{\tilde{\tau} ln(\tilde{\tau})}\left(\tau^{-1}\right)$ \\

  \hline
  $3<\gamma<5$ & $\tilde{\tau}^{\frac{1}{2}}\left(\tau^{\frac{1}{\gamma-3}}\right)$ \vline
 $ \ \ \tilde{H}^{\frac{8-\gamma}{9}}$ & $
\tilde{\tau}^{\frac{\gamma-5}{2}}\left(\tau^{\frac{5-\gamma}{\gamma-3}}\right)$
&
$\tilde{\tau}^{\frac{3-\gamma}{2}}\left(\tau^{-1}\right)$ \\

  \hline
\end{tabular}
\end{center}

\caption{Comparison of the critical behavior of Ising model on
scale free random networks \cite{dgm} (written inside parenthesis)
and its edge-dual network.}\label{tab}

\end{table}

\section{Conclusion}\label{6}
In summary we studied the Ising model with nearest neighbor
interactions on the edge-dual of uncorrelated random networks. We
stated a simple relation between the partition function of this
model and that of an Ising model with next nearest neighbor
interactions on a tree-like network. High and low temperature
expansions of the partition function were also derived. As a
simple example we studied the Ising model on the edge-dual of
Bethe lattices using the well known recurrence relation procedure.
We finally generalized this study to the edge-dual of uncorrelated
random networks. Although the critical temperature of Ising model
on edge-dual network is higher than the one in the random network,
both quantities become infinite in the same point, that is when
the second moment of the degree distribution of random network,
$<k^2>$, becomes infinite. This fact reflects the robustness of
edge-dual networks against thermal fluctuations, a property which
can be attributed to the large number of triangles and the special
structure of the edge-dual networks. We also derived the critical
behavior of Ising model on edge-dual network of an uncorrelated
random scale free network. The results show that this behavior is
significantly different from the one seen in the uncorrelated
random networks.

\subsection*{Acknowledgment}

The author is grateful to V. Karimipour for helpful discussions
and useful suggestions.

\end{document}